\documentstyle[sprocl]{article}

\def\fr#1#2{{{#1} \over {#2}}}
\def\gsim{\mathrel{\rlap{\lower4pt\hbox{\hskip1pt$\sim$}}
    \raise1pt\hbox{$>$}}}

\def\be{\begin{equation}}
\def\ee{\end{equation}}
\def\bea{\begin{eqnarray}}
\def\eea{\end{eqnarray}}

\begin{document}

\title{SOME CONSIDERATIONS REGARDING LORENTZ-VIOLATING THEORIES}

\author{RALF LEHNERT}

\address{Physics Department, Indiana University\\
Bloomington, IN 47405, U.S.A.
\\E-mail: rlehnert@indiana.edu} 

\maketitle\abstracts{
We investigate the compatibility
of Lorentz-violating quantum field theories
with the requirements
of causality and stability.
A general renormalizable model
for free massive fermions
indicates that these requirements are satisfied
at low energies provided
the couplings controlling the breaking are small.
However, for high energies
either microcausality or energy positivity
or both are violated in some observer frame.
We find evidence that this difficulty can be avoided
if the model is interpreted
as a sub-Planckian approximation
originating from a nonlocal theory
with spontaneous Lorentz violation.
The present study thereby supports the validity
of the standard-model extension
as the low-energy limit
of any realistic string theory
that exhibits spontaneous Lorentz breaking.}

\section{Introduction}

\noindent 
From a theoretical point of view,
the minimal SU(3)$\times$SU(2)$\times$U(1) standard model
leaves unresolved a variety of issues.
It is therefore believed to be the low-energy limit
of an underlying framework
that also includes a quantum description of gravity.
On the other hand,
the standard model is phenomenologically successful,
so observable effects
from the presumed underlying physics
must be minuscule.
It then becomes an interesting challenge
to identify possible experimental signals
from such a fundamental theory
accessible with present techniques.

A candidate signal of this type
is the violation CPT and Lorentz invariance:
In conventional renormalizable
gauge theories including the standard model,
these two symmetries are linked
by CPT theorem\cite{pauli} and hold exactly.
In contrast,
attempts to construct an underlying framework
often involve ingredients
that bypass the CPT theorem.
For example,
string (M) theory is known to admit
spontaneous Lorentz and CPT violation.\cite{str}
Other frameworks can also lead to similar
low-energy effects.\cite{klink,ch}

For the microscopic description
of possible observable signals at presently accessible energy scales,
an extension of the minimal standard model of particle physics
has been developed.\cite{ck}
This standard-model extension
has provided the basis for numerous experimental
investigations discussed during this meeting and elsewhere,
which  constrain CPT and Lorentz violation.\cite{exp}

In this talk, we study the fundamental properties
of causality and stability
in the context of the Lorentz-violating standard-model extension.
These two properties appear essential for realistic theories,
for it would be difficult to make meaningful experimental predictions
without either causality or stability.
In particular,
it is of interest whether these two requirements
constrain the parameter space and the range of validity
of the standard-model extension,
and whether insight into the underlying theory can be gained.
Although the calculations presented here are carried out
for free massive fermions,
we expect that most of our results can be straightforwardly generalized
to the other sectors of the standard-model extension.

\section{Framework}

\noindent
The general Lorentz-violating Lagrangian
for a single spin-$\frac{1}{2}$ fermion
\cite{ck}
can be cast into a variety of forms.
One such form
reminiscent of the ordinary Dirac Lagrangian
and emphasizing the derivative structure
is \cite{kl99} 
\begin{equation}
{\cal L} = \frac{1}{2}{\it i}\overline{\psi}
{\Gamma}^{\nu}\hspace{-.15cm}
\stackrel{\;\leftrightarrow}
{\partial}_{\!\nu}\hspace{-.1cm}{\psi}
-\overline{\psi}M{\psi}
\label{lagr}
\quad ,
\end{equation}
where

\begin{equation}
{\Gamma}^{\nu}:={\gamma}^{\nu}+c^{\mu \nu}
{\gamma}_{\mu}+d^{\mu \nu}{\gamma}_{5}
{\gamma}_{\mu}+e^{\nu}+if^{\nu}{\gamma}_{5}
+\frac{1}{2}g^{\lambda \mu \nu}
{\sigma}_{\lambda \mu}
\label{Gam}
\quad ,
\end{equation}
and

\begin{equation}
M:=m+a_{\mu}{\gamma}^{\mu}+b_{\mu}{\gamma}_{5}
{\gamma}^{\mu}+\frac{1}{2}H^{\mu \nu}
{\sigma}_{\mu \nu}
\label{M}
\quad .
\end{equation}
The gamma matrices $\{1, \gamma_5,\gamma^{\mu},
\gamma_5\gamma^{\mu}, \sigma^{\mu \nu}\}$
have conventional properties,
and the signature of the Minkowski metric
$\eta_{\mu\nu}$ is $-2$.
The extent of Lorentz violation
is described by the parameters
$a_{\mu}$, $b_{\mu}$, $c_{\mu \nu}$,
$d_{\mu \nu}$, $e_{\mu}$, $f_{\mu}$,
$g_{\mu \nu \lambda}$ and $H_{\mu \nu}$.
As a consequence
of the presumed hermiticity
of the Lagrangian,
all these coefficients
are real,
with
$c_{\mu \nu}$ and $d_{\mu \nu}$ traceless,
$g_{\mu \nu \lambda}$ antisymmetric
in its first two indices
and $H_{\mu \nu}$ antisymmetric.
While all the parameters
violate Lorentz invariance,
only
$a_{\mu}$, $b_{\mu}$, $e_{\mu}$, $f_{\mu}$
and $g_{\mu \nu \lambda}$
break CPT symmetry as well.

Since no departures from Lorentz symmetry
have been observed to date,
all Lorentz-breaking parameters
must be minuscule in a certain class
of observer inertial frames
called {\it concordant frames},
and the Earth must move nonrelativistically
with respect to these frames.
Throughout this talk,
we shall work under the assumption
that the size of the Lorentz violation
is such that,
in a concordant frame,
a hermitian hamiltonian can be found
and the dispersion relation still exhibits
two positive- and two negative-valued roots,\cite{kl01,kljmp}
paralleling the conventional Dirac case.
The lagrangian can then be canonically quantized
such that the energy is positive definite.\cite{kl01}

\section{Microcausality and Stability}

\noindent
A quantum field theory is microcausal
if any two local observables
with spacelike separation
can be measured independently.
This is guaranteed
if any two local,
spatially separated operators
commute.
In the present case,
such local operators
are fermion bilinears
and the above condition
is satisfied if
\begin{equation}
iS(x-x^{\prime})=\{\psi(x),
\overline{\psi}(x^{\prime})\}=0
\quad,
\qquad(x-x^{\prime})^2<0
\label{anticom}
\end{equation}
holds.
Note that the anticommutator
function $S(x-x^{\prime})$ only depends
on the coordinate differences
due to translational invariance.

To determine
an integral representation
for the $S(z)$,
we insert the plane-wave expansion
of the field operators\cite{kl01}
into the anticommutator and use the generalization
of the conventional spinor projectors.\cite{kljmp}
This gives the following expression:
\begin{equation}
S(z)=\int_{C}\fr{d^4\lambda}{(2\pi)^4}
e^{-i\lambda\cdot z}
\fr{{\rm cof}(\Gamma_{\mu}\lambda^{\mu}-M)}
{\det(\Gamma_{\mu}\lambda^{\mu}-M)}
\label{pwexp2}
\quad,
\end{equation}
where $C$ is the usual contour
encircling all poles in clockwise direction,
and ${\rm cof}(\cdot)$ denotes the matrix of cofactors.
Notice that $\lambda^{\mu}$
can be replaced by $i\partial^{\mu}$ in the numerator of the integrand.
It is then possible to pull the cofactor matrix outside the integral,
because the contour $C$ can be deformed
such that the integrand is analytic in a neighborhood of $C$.\cite{sg}
We obtain for the anticommutator function
\begin{equation}
S(z)=
{\rm cof}(\Gamma^{\mu} i\partial_{\mu}-M)
\int_{C} \fr{d^4\lambda} {(2\pi)^4}
\fr{e^{-i\lambda\cdot z}} {\det(\Gamma^{\mu}\lambda_{\mu}-M)}
\quad.
\label{ffgreen}
\end{equation}

Next, we study $S(z)$ outside the lightcone.
We can take advantage
of observer Lorentz invariance
and boost to a frame such that
$z^{\mu}=(0,\vec{z})$.
To make further progress,
it is necessary to investigate the pole structure
of the integrand.
Due to the above observer transformation
we may no longer assume to be working in a concordant frame.
In particular,
it may not be possible to find a hermitian hamiltonian,
so that complex eigenenergies may occur.
Since the eigenenergies determine the location of the poles
of the integrand, the contour $C$ may fail to encircle them all.
Thus, the case where a hermitian hamiltonian
(and therefore real eigenenergies)
exist in all frames
has to be distinguished.
We consider this case first.

A sufficient condition for the hermiticity of the hamiltonian
in all observer frames is that the derivative structure
of lagrangian (\ref{lagr}) is the conventional one,
{\it i.e.,} $\Gamma^{\mu}=\gamma^{\mu}$.
Then, all four roots
$E_{(j)}(\vec{\lambda})$, $j=1,\ldots,4$,
of the dispersion relation appearing in the denominator
of the integrand in Eq.\ (\ref{ffgreen})
are real.
In this case,
the contour integration can be directly performed.\cite{kl01}
This argument confirms microcausality
for the case $\Gamma^{\mu}=\gamma^{\mu}$.

In cases
when there exist observer frames
that fail to admit the definition
of a hermitian hamiltonian,
the above line of reasoning cannot be employed,
and microcausality may break down.
For example,
consider a model with $c_{00}$ parameter only.
The anticommutator function for this model
is given explicitly by
\begin{equation}
S(z)=(i\zeta\gamma^0\partial^0
-i\gamma^j\partial^j+m)
\fr{1}{4\pi\zeta r}
\fr{\partial}{\partial r}[\Theta(w^2) J_0(m\sqrt{w^2})]
\quad,
\label{prop}
\end{equation}
where
$\zeta=1+c_{00}$,
$r=|\vec{z}|$,
$w^2=(z^0/\zeta)^2-\vec{z}^2$,
$\Theta$ denotes the Heaviside step function
and $J_0(y)$ is the zeroth-order Bessel function.
It follows
that the anticommutator function $S(z)$
vanishes only in the region defined by
$z^0<(1+c_{00})|\vec{z}|$.
The propagation of signals therefore could occur with maximal speed
$1/(1+c_{00})$.
For negative values of $c_{00}$,
this exceeds 1 and hence violates microcausality.

The question arises,
at which energy scale
this breakdown of microcausility occurs.
To this end,
it is useful to introduce a definition
of the velocity of a particle valid for arbitrary 3-momenta.
Even in the conventional Lorentz- and CPT-symmetric case,
the notion of a quantum velocity operator is nontrivial.
The issue is further complicated in the present context.\cite{ck}
Here, we consider the group velocity
defined for a monochromatic wave
in terms of the dispersion relation.
This choice seems appropriate for a variety of reasons.\cite{kl01}
Insight about the scale $\tilde{M}$ of microcausality violations
can then be gained by determining the value of the 3-momentum
at which the the group velocity reaches 1.
Analyses for a variety of parameter combinations
yield
\begin{equation}
\tilde{M}\gsim {\cal O}(M_P)
\quad .
\label{mscale1}
\end{equation}
Here, we have assumed that the parameters
$c_{\mu\nu}$, $d_{\mu\nu}$, $e_{\mu}$,
$f_{\mu}$ and $g_{\mu\nu\lambda}$
are of order $m/M_P$,
where $M_P$ denotes the Planck scale.

We mention in passing
that the conclusion of microcausality breakdown
at ${\cal O}(M_P)$ may be invalid
if the $c_{\mu\nu}$ coefficient is nonzero.
For example,
in the above model with only a coupling $c_{00}<0$,
one can show that $\tilde{M}\gsim{\cal O}(\sqrt{mM_P})$. 
The effect of the $c_{\mu\nu}$ parameter on the dispersion relation
is special for the following reason:
The general spinorial and derivative structure
of the associated quadratic field term
is identical to the conventional Dirac kinetic term.
Thus, it is a first-order correction
to an existing zeroth-order term.
None of the other Lorentz-violating couplings
exhibits this feature.

The above analysis shows
that the standard-model extension
can develop problems when the symmetry-breaking scale
is approached.
This should not come as a surprise
because the effects of the presumed underlying theory
are likely to be no longer negligible at these energies.
However,
given the impracticality
of achieving Planck-scale momenta
in the laboratory,
the issue of microcausality breakdown
is largely unimportant at the level of the standard-model extension.

Another important ingredient
for realistic field theories is the requirement of stability.
A field theory is stable if the energy is positive definite
{\it in all observer frames}.
This implies that the 4-momenta of all one-particle states
in a particular frame must be timelike or lightlike
with nonnegative 0th component.
Only under this last condition,
does energy positivity become an observer-invariant notion.
This is satisfied in the conventional Dirac case.

In the present context,
the energy is positive definite in concordant frames.
In these frames,
the dispersion relation
has still two positive- and two negative-valued roots,
which yield positive particle energies
after the usual reinterpreta-tion.\cite{kl01}
However,
contrary to the conventional case,
these energies are in some instances
0th components of spacelike 4-momenta.
As a result,
energy positivity becomes observer-dependent.

As an example,
consider a model that has all Lorentz-violating
parameters except $b_{\mu}$ set to zero.
The dispersion relation for this model is given by
\begin{equation}
(\lambda^2-b^2-m^2)^2+4b^2\lambda^2 -4(b\cdot\lambda)^2=0
\quad .
\label{bdisp}
\end{equation}
It is straightforward to show
that observer frames in which
$b_{\mu}=(b_0,0,0,b_3)$ and ${b_3}^2>m^2+|b^{\mu}b_{\mu}|$
can always be chosen.
In such a frame,
the spacelike 4-vectors
${\lambda^{\mu}}_{\pm}=(0,0,0,p_{\pm})$
satisfy the dispersion relation (\ref{bdisp}).
Here, the real quantities $p_{\pm}$
are defined by
\begin{equation}
{p_{\pm}}^2=(2{b_3}^2+b^2-m^2)
\pm\sqrt{(2{b_3}^2+b^2-m^2)^2-(m^2+b^2)^2}
\quad .
\label{moment}
\end{equation}
Moreover, the existence of these spacelike solution
remains unaffected,
when a nonzero $a_{\mu}$ coefficient is included.

The instabilities resulting from these spacelike
4-momenta are most transparent
for sufficiently boosted observers:
It is always possible to convert a spacelike vector
with a positive 0th component to one with a negative 0th component
by an appropriate observer Lorentz transformation.
In the present case,
this means that there exist otherwise acceptable observer frames
in which a single root of the dispersion relation 
involves both positive and negative energies
for varying 3-momenta.
In such observer frames,
the canonical quantization procedure fails.

In concordant frames, the energy is positive definite.
However, the physics is independent of the observer,
so the appearance of negative energies in a boosted frame
must also lead to instabilities in the concordant frames.
The above discussions implies that these instabilities
can only be associated with the spacelike momenta
satisfying the dispersion relation.
To illustrate this,
let us introduce a U(1) gauge interaction for the moment
because in the free fermion model
the particle number is conserved.
As an example,
consider the following process in a concordant frame:
A high-energy fermion emits a virtual photon,
which then decays into a fermion-antifermion pair.
We can write this as
\begin{equation}
f_{+1}\longrightarrow f_{+1} +f_{+1}+\bar{f}_{-1}
\quad ,
\label{decay}
\end{equation}
where $f$ and $\bar{f}$ denote fermions and antifermions,
respectively,
and the subscript labels the helicity state.
In ordinary QED,
such a process is kinematically forbidden
even though both the U(1) charge and angular momentum are conserved.
However it can occur in the present context
if the incoming fermion
has an appropriate spacelike 4-momentum.\cite{kl01}
Thus, there exist unstable single-particle states.

The scale $\tilde{M}$ of the 3-momentum
at which spacelike 4-momenta occur
can be calculated explicitly
for various parameter combination.
We find that
\begin{equation}
\tilde{M} \gsim {\cal O}(M_P)
\quad ,
\label{bscale}
\end{equation}
where we have assumed that the derivative-coupling coefficients
have the same suppression as in the microcausality case,
and the remaining parameters $a_{\mu}$, $b_{\mu}$ and $H_{\mu\nu}$
are of order $m^2/M_P$.
This estimate shows that the instabilities
appear only for Planck-scale 4-momenta in a concordant frame.
The corresponding negative energies 
occur only for Planck-boosted observers
relative to this frame.
Since the Earth moves nonrelativistically with
respect to concordant frames,
the model maintains stability
for all experimentally attainable physical momenta
and in all experimentally attainable observer frames.

As for microcausality,
the presence of a $c_{\mu\nu}$ parameter
can invalidate (\ref{bscale}).
For example,
a model with a positive $c_{00}$ coefficient only,
exhibits spacelike momenta at a scale of order
$\sqrt{mM_P}$.
It follows that when microcausality and stability
are imposed on a model with a $c_{\mu\nu}$ coupling,
effects from the presumed underlying theory
are likely to become non-negligible
already at energies close to the geometric mean of $M_P$ and $m$.
As this scale is within reach of some experiments,
a theoretical analysis of such effects
may require high-energy corrections to the standard-model extension.
In the next section,
we discuss a possible type of such corrections.

\section{High-Energy Effects}

\noindent
The results from the previous section indicate
that quantum field theories of massive fermions
containing terms explicitly breaking Lorentz invariance
can develop difficulties with mircocausality or stability.
However, in concordant frames,
these difficulties primarily appear
as the Planck scale is approached.
The question arises,
whether there exist combinations of Lorentz-violating coefficients
that maintain both causality and stability.
Many parameter combinations
eliminate one of the two difficulties.
However,
we are unaware of any set of values of the couplings
$a_{\mu}$, $b_\mu$, $\ldots$, $H_{\mu\nu}$
that simultaneously guarantee microcausality and stability
at all energy scales.
Moreover,
it has been shown rigorously
that in {\it conventional} quantum field theory,
such a set of parameters would have to yield
the ordinary Lorentz-symmetric dispersion relation.\cite{be}
It is then likely that these Lorentz-breaking parameters
can be absorbed into a field redefinition
and remain unobservable.

The Lorentz-violating standard-model extension
was developed following a top-down approach.
The original motivation was the possibility
of spontaneous Lorentz-symmetry breakdown
in an underlying framework such as strings.\cite{str}
Indeed, the standard-model extension includes
all Lorentz-violating, but observer-invariant, terms
compatible with renormalizability and the usual gauge structure.
It is thus the low-energy limit
of any potential spontaneous Lorentz breaking
in a more fundamental theory.
It is therefore not surprising
that difficulties develop as the Planck-scale is approached.
One would expect higher-order nonrenormalizable operators
to gain importance.
On the other hand,
the essential status of the requirements of causality and stability
suggests to adopt the inverse line of reasoning.
Such a bottom-up approach could provide
valuable insights into the nature of the underlying theory
at the Planck scale.

The standard-model extension breaks Lorentz invariance explicitly.
However, a desirable feature of the fundamental theory
is spontaneous symmetry breaking.
One immediate advantage of this mechanism is that the dynamics
remains Lorentz covariant.
Therefore it does not come as a surprise
that such an underlying framework avoids at least some of  
the difficulties plaguing more general models 
involving Lorentz and CPT violation.
For instance,
one consequence of the spontaneous character
of the Lorentz violation is
that observer invariance is naturally maintained.
In the previous section,
this property has proved to be an important advantage.
In contrast,
if observer Lorentz invariance is imposed
in a theory with explicit Lorentz breaking,
an additional \it ad hoc \rm choice is required.

Another effect of spontaneous Lorentz violation is
that the parameters
$a_{\mu}$, $b_\mu$, $\ldots$, $H_{\mu\nu}$
are only fixed at low energies.
As the Planck scale is approached,
they must be associated with dynamical fields.
A natural question is,
whether these fluctuations alone can simultaneously maintain
microcausality and stability.
This issue has been previously been discussed
in the context of a toy model.
It was shown that a satisfactory resolution
within the context of ordinary point-particle field theory
seems unlikely.\cite{kl01}
This is consistent with other ideas.\cite{be}
As expected,
ingredients beyond conventional quantum field theory
appear necessary.

A class of theories with free-field terms
maintaining causality and stability
must contain terms beyond the ones in Eq.\ (\ref{lagr}).
The new terms have to be nonrenormalizable,
and in a realistic scenario with spontaneous Lorentz violation 
they would correspond to higher-order nonrenormalizable operators 
correcting the standard-model extension at energies
determined by the Planck scale.

The first step is to investigate
whether any type of dispersion relation
can satisfy all the requirements for consistency.
In a concordant frame,
such a dispersion relation
would reproduce the physics of Eq.\ (\ref{lagr})
for small 3-momenta
but would avoid group velocities exceeding 1 and spacelike 4-momenta
for large 3-momenta.
These requirements could be implemented 
by combining the Lorentz- and CPT-breaking parameters
with a suitable factor suppressing them only at large 3-momenta.
This factor must be essentially constant at small 3-momenta
and must overwhelm polynomial powers at large 3-momenta. 
Since the the size of 3-momenta
is frame dependent,
it is to be expected that a suitable factor
would also be frame-dependent
and hence involve Lorentz- and CPT-violating coefficients.

To make further progress,
it is useful to consider explicit examples.
To simplify the discussion,
the masses and the Lorentz-violating parameters
are taken to be of order 1
in appropriate units.
This makes it possible to focus on resolving
the problems of stability and causality
at Planck-scale energies in a concordant frame
without the complications introduced by the hierarchy of scales.

Consider a model with a negative $c_{00}$ parameter only.
As discussed in the previous section,
this model violates microcausality at high energies.
The replacement
$c_{00}\rightarrow c_{00}\exp (c_{00}{\lambda_0}^2)$
in the dispersion relation
has been shown to result in subluminal group velocities
for all 3-momenta
without introducing instabilities.\cite{kl01}
In an arbitrary frame,
this modification takes the form
\begin{equation}
c_{\mu\nu}\rightarrow c_{\mu\nu}
\exp (c_{\mu\nu}\lambda^{\mu}\lambda^{\nu})
\quad ,
\label{crepl}
\end{equation}
establishing observer invariance
of the resulting dispersion relation.
It can also be shown
that introducing similar exponential suppression factors
in models with instabilities
can also resolve this problem
while maintaining subluminal group velocities.

The above demonstrations prove that stable and causal
dispersion relations 
violating Lorentz and CPT symmetry can exist.
The occurrence of transcendental functions of the 4-momenta
corresponds to derivative couplings of arbitrary order
in the lagrangian.
A satisfactory framework incorporating Lorentz and CPT violation
appears necessarily to be nonlocal in this sense.
Although it is in principle conceivable that
a model with explicit Lorentz breaking might satisfy the requirements
of causality and stability,
it would appear somewhat contrived to
implement both the necessary observer Lorentz invariance
and nonlocal couplings by hand.
On the other hand,
one can  see that spontaneous Lorentz and CPT violation
in a nonlocal theory 
can naturally yield the desired ingredients
for stability and causality at all scales.

It would be interesting to identify theories 
from which these dispersion relations emerge naturally.
A promising candidate for this type of framework is string theory.
It provided the original motivation for the construction
of the standard-model extension.
Moreover, strings are known to admit spontaneous Lorentz
violation and they have nonlocal interaction.
A complete treatment of this question would be desirable,
but is hampered by the absence
of a satisfactory realistic string theory.
Instead, we consider the field theory
of the open bosonic string as an example
and show that its structure is compatible
with dispersion relations of the desired type.

The open bosonic string has no fermion modes.
We will therefore consider the scalar tachyon.
The relevant quadratic terms of the lagrangian for the tachyon
in the presence of Lorentz violation are given by:\cite{str}
\begin{eqnarray}
{\cal L} &\supset &
\fr{1}{2} \partial_\mu \phi \partial^\mu \phi  
+ (\alpha^{\prime -1} + k_0) \phi^2
+ \ldots
+ k_1 \langle B_{\mu\nu} \rangle \partial^\mu \phi \partial^\nu \phi  
\nonumber\\
&&\qquad
+ \ldots + k_2 \langle D_{\mu\nu\rho\sigma} \rangle
\partial^\mu \phi \partial^\nu \phi \partial^\rho \phi
\partial^\sigma \phi  
+ \ldots
\quad .
\label{string}
\end{eqnarray}
Here,
the scalar parameters $k_0$, $k_1$, $k_2$, $\ldots$
are determined by the theory,
but their specific values are irrelevant 
for the present considerations.
Each ellipsis represents quadratic terms involving
vacuum expectation values of other tensors
and terms with powers of $\lambda^2$.

For a plane-wave tachyon solution,
the structure of the dispersion relation
resulting from lagrangian (\ref{string})
indeed exhibits features
needed to maintain causality and stability.
For example,
it contains all terms of the dispersion relation that results
from the replacement (\ref{crepl}) in the $c_{00}$ model,
as the reader is invited to verify.

We emphasize that the purpose of the above discussion
is only to provide an outline indicating how
a satisfactory dispersion relation for Lorentz violation
could emerge in the context of string theory.
In particular,
we do not claim that the tachyon itself
must \it necessarily \rm obey such a relation,
although it is conceivable that it does.\cite{str}
Here,
the tachyon dispersion relation is used 
merely as an illustration to display explicitly
the appearance of nonlocal couplings in string theory 
that could be appropriate for 
a stable and causal theory with spontaneous Lorentz violation.
This type of coupling is generic both for other fields in
the open bosonic string 
and for fields in other string theories,
including ones with fermions.

\end{document}